# A Miniaturized Design of Drift Tube Ion Mobility Spectrometry

**Ziyi FangF**[a] **and Mang He**[a,b]

[a] *School of Integrated Circuits and Electronics,*
  *Beijing Institute of Technology, Beijing, 100081, China*
[b] *Senior Member, IEEE*
  E-*mail*: hemang@bit.edu.cn

ABSTRACT: The drift tube ion mobility spectrometry (DTIMS) is a device used to detect trace chemicals, and its miniaturized design is always desired in practice engineering. In previous studies, little attention has been paid to the reaction region of the DTIMS in the miniaturization design. In this letter, a series of electrodes with uniform interval is introduced in the reaction region to achieve miniaturization of the device. It is found that the added electrodes enlarge the effective coverage area of the ion swarms in front of the TP shutter and increase the concentration of ions entering the drift region within the pulse period, which can effectively reduce the overall size of the device. In addition, the effect of the thickness-to-interval (T/I) ratio of electrodes is investigated and optimized, and the resolution of the DTIMS is significantly improved with appropriate T/I ratio despite its compact size. The prototype of the device is fabricated, and the size of the DTIMS is reduced by about 50%, while the measured resolution is improved by 80% as compared with the existing commercial product.

# Contents

## 1. Introduction

The drift tube ion mobility spectrometry (DTIMS), whose structure is shown in Fig.1, is used for trace detection of some volatile organic compounds, such as drugs, explosives, chemical warfare agents and atmospheric pollutants. The device usually works at ambient pressure, and the gas-phase ions of different compounds move in the electric field with their specific averaged drift velocities from which the DTIMS can discriminate the type of compound. In the past decades, the trend of DTIMS design has been focused on miniaturization and resolution improvement [1], and several approaches have been proposed to achieve this goal, such as optimizing the ion motion under the complex electric and flow fields in the drift region [2]-[10]. On the other hand, the reaction region that controls the ion forming process from the ion source to the TP shutter and then to the drift region, also plays an important role in the optimal design and size reduction of the DTIMS. However, little attention has been paid to the design of the reaction region in previous studies, and its effects on the performance of DTIMS has not been fully taken into account yet.

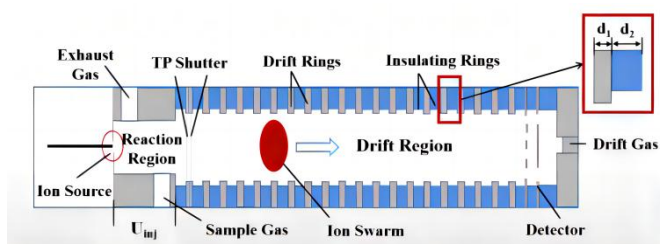

**Figure 1**. The structure of DTIMS.

In this letter, a multi-physics model is set up to simulate the entire working process of the DTIMS from the ion source to the TP shutter in the reaction region and then ion swarms motion in the drift region, which provides an effective tool to optimize the ion distributions in the reaction region and the coverage area of ion swarm in front of the TP shutter. Based on the numerical simulations, the appropriate injection voltage is determined, and then a miniaturized design of DTIMS is realized by adding a series of uniformly placed electrodes in the reaction region. Furthermore, the thickness-to-interval (T/I) ratio of the electrodes is optimized to improve the resolution of the DTIMS with compact size. A prototype of the new design is fabricated and measured to validate the effectiveness of the proposed method.

## 2. Miniaturized design of the DTIMS

### 2.1 Working principle of the DTIMS and multi-physics modeling

As shown in Fig. 1, initially, the sample gases enter the reaction region and are ionized into ion swarms by the ion source. Under an injection voltage $U_{inj}$, the ion swarms are injected into the drift region, usually a cylindrical tube, through the TP shutter that is opened periodically. At ambient pressure, these ion swarms move through the purified air in the electric field produced by the drift rings (electrodes) with constant neighboring voltage. The drift gas provides a purified and constant air flow for collision-based movement of the ion swarms, and a weak current is formed after the ion swarms reaching the detector. Due to specific mobility of each type of ions, the drift time of different ions from the TP shutter to the detector is distinct. Therefore, the response of detector versus drift time (mobility spectrum) can be used to distinguish the kind of ions.

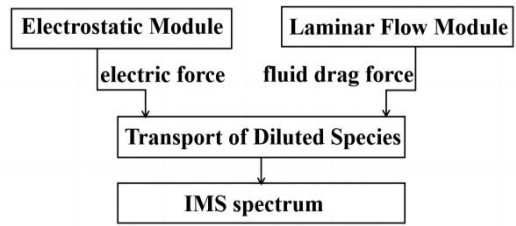

**Figure 2.** Flowchart of Multi-physics Modeling.

The entire multi-physics working process of the DTIMS can be numerically simulated by using the Electrostatic module, the Laminar Flow module, and the Transport of Diluted Species module of the commercial software COMSOL in a combined way as shown in the flowchart (Fig.2). As a numerical example, Fig.3 shows the simulated normalized detector currents of two substances with different mobilities, which demonstrates that the multi-physics model can well distinguish the type of substance. In Fig.3, the mobility K of substance is defined as the ratio of its motion velocity v to the intensity of the applied electric field E., i.e., K=v/E.

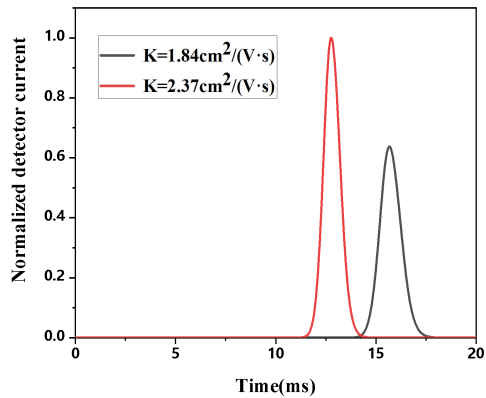

**Figure 3.** Normalized detector current for two substances with different K.

Due to the random collision between the ion swarms and the neutral particles in the drift region, even for ions of the same type, the arrival time of each ion at the detector is not precisely same. Therefore, the current response of the detector versus drift time approximately obeys the Gaussian's distribution [1]. The resolution $R_p$ can be defined as the primary parameter to evaluate the ability of the DTIMS to discriminate different types of ions, which reads:

$$R_p = t_d / w_h \quad (1)$$

where $t_d$ and $w_h$ are the center and the half-peak width of the Gaussian pulse, respectively.

## 2.2 Miniaturization of the DTIMS

The design starts from an existing commercial product TR1000DC -C Handheld Explosives and Narcotics Trace Detector (Courtesy of Nuctech Co., Ltd). Through numerical simulations, we observe that the injection voltage $U_{inj}$ plays a critical role in affecting the uniformity of the electric field in the reaction region, and further the ion swarm motion in the drift region, and finally the resolution of the DTIMS. Fig.4 shows the simulated distribution of ion swarms in the reaction region. It can be seen that the smaller the $U_{inj}$ is, the larger the coverage area of ion swarms gathered in front of the TP shutter.

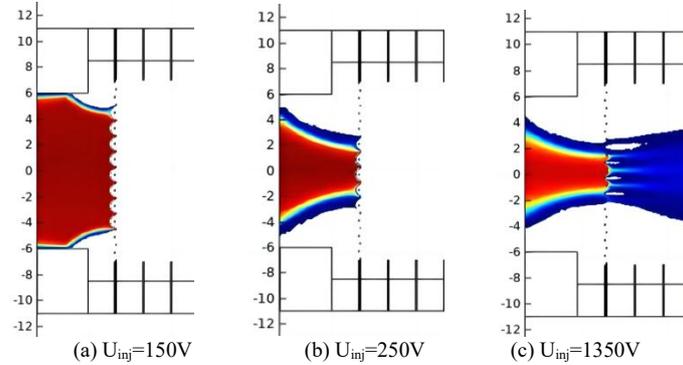

(a) $U_{inj}$=150V      (b) $U_{inj}$=250V      (c) $U_{inj}$=1350V

**Figure 4.** The simulated distribution of ions in the reaction region under various $U_{inj}$.

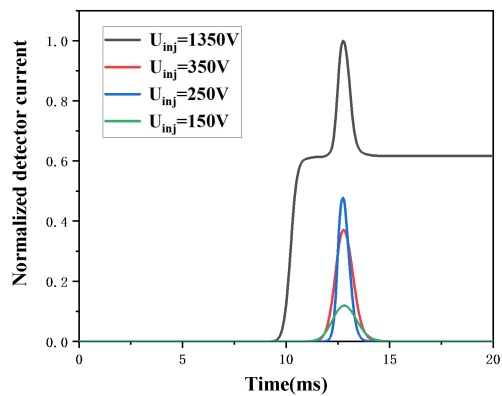

**Figure 5.** Normalized detector current under various $U_{inj}$.

When $U_{inj}$ is beyond 250 V, the TP shutter cannot be closed tightly any longer as shown in Fig.4(c), which results in a sharply decreased resolution in Fig.5 (black line and red line). On the other hand, when the $U_{inj}$ is lower than 150 V, injection of the ion swarms into the drift region is slow, leading to a poor sensitivity as well (green line in Fig. 5). Hence, the appropriate value of $U_{inj}$ (250 V or so in this case) should be determined with great care by numerical simulations.

The simulated ions' distribution in the reaction region is shown in Fig. 6(a), and the diameter of the ion swarm coverage area in front of the TP shutter is about 6 mm. While in Fig. 6(b), in order to reduce the size of the DTIMS, the inner diameters of the reaction and drift regions of the original structure are reduced from 12 mm to 9 mm and from 17 mm to 11 mm, respectively, it is found that coverage diameter decreases to 3 mm. However, as seen in Fig. 6 (b) and Fig. 7, the reduced coverage area in front of the TP shutter allows fewer ion swarms to enter the drift region and then lowers the resolution of the device (blue line) as compared with that of the original design (black line).

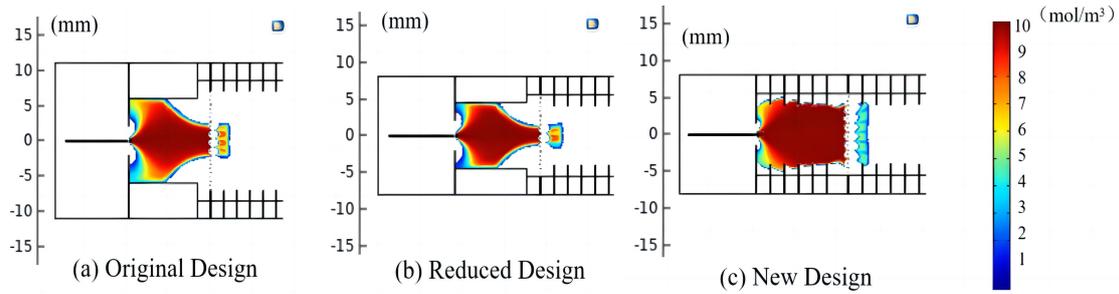

Figure 6. Evolution of the miniaturized designs and simulated ion distributions in the reaction regions.

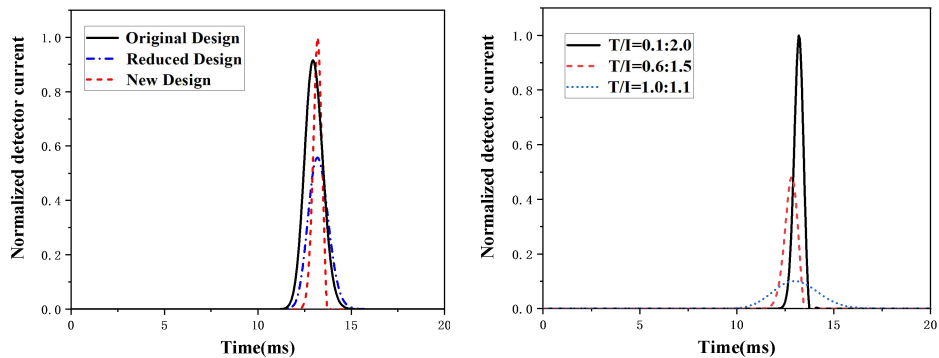

Figure 7. Normalized detector responses of the three designs in Fig.6 versus drift time. (left)
Figure 8. Normalized detector responses for different T/I ratio.(right)

To recover or even improve the resolution of the DTIMS with compact-size design in Fig. 6(b), we introduce a few electrodes with uniform intervals in the reaction region as shown in Fig. 6(c). It is interesting to find that the distribution of ions is more uniform and the coverage area in front of the TP shutter is significantly enlarged, so more ion swarms can be injected into the drift region. As a result, the resolution of the new design (red line) is much better than that of the design in Fig. 6(b) and even higher than that of the original structure.

Meanwhile, we find that the T/I ratio of the electrodes, i.e., $d_1/d_2$ as shown in Fig. 1, has critical effects on the resolution of the device. It is seen in Fig. 8 that the resolution of DTIMS is rather sensitive to the T/I ratio, and the resolution increases sharply as the T/I ratio reduces. Nevertheless, excessively thin drift rings are hard to manufacture and assemble, thus we set the optimal values of $d_1$ and $d_2$ to 0.1 mm and 2.0 mm, respectively.

## 3. Measurement

The proposed design is fabricated, and the comparison of the drift tube of the new prototype and the original structure is shown in Fig. 9 to illustrate the effectiveness of size reduction. In Fig. 10, the measured current response of the detector versus ion swarms' drift time of both the new and original designs are displayed, and the normalized response current intensity is shown in Fig. 11. From the measured data, the resolutions of the original and new designs can be calculated as 23.6 and 42.3, respectively. Apparently, the resolution of the device is improved by 80%, even the size of new design is only about 50% of the existing product.

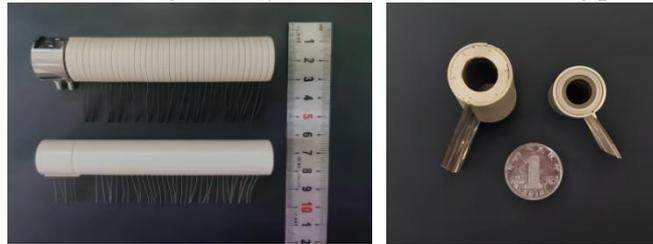

**Figure 9.** Fabricated drift tubes (a) top view, (b) axial view.

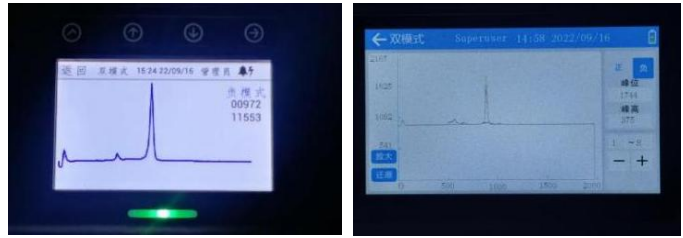

**Figure 10.** Measured detector responses: (a) original design; (b) new design.

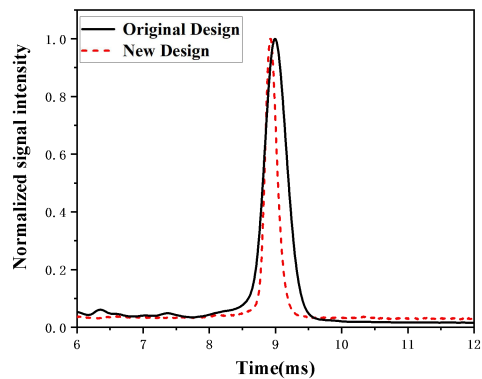

**Figure 11**. Comparison of the measured current intensities on detector after normalizing.

## 4. Conclusion

In this letter, a miniaturized design of DTIMS is realized by introducing a series of electrodes with uniform interval in the reaction region. The thickness-to-interval ratio of the added electrode is optimized to achieve a high resolution of the device. The size of the miniaturized design is reduced by 50% as compared with the existing commercial product, while the resolution is 80% higher. The prototype of the new design is fabricated, and the measured data demonstrate the effectiveness of the proposed method.